\documentclass[fleqn,usenatbib]{mnras}

\usepackage{newtxtext}

\usepackage[T1]{fontenc}

\DeclareRobustCommand{\VAN}[3]{#2}
\let\VANthebibliography\thebibliography
\def\thebibliography{\DeclareRobustCommand{\VAN}[3]{##3}\VANthebibliography}

\usepackage{graphicx}	
\usepackage{amsmath}	
\usepackage{amssymb}	

\title[Short title, max. 45 characters]{Observations of two super fast rotator NEAs: 2021 NY$_1$ and 2022 AB}

\author[J. Licandro et al.]{
Javier Licandro$^{1,2}$,\thanks{E-mail: jlicandr@iac.es}
Marcel Popescu$^{3,4}$,
Eri Tatsumi$^{1,2,5}$,
Miguel R. Alarcon$^{1,2}$,
Miquel Serra-Ricart$^{1,2}$,
\newauthor
Hissa Medeiros$^{1,2}$,
David Morate$^{1,2}$,
Fernando Tinaut-Ruano$^{1,2}$,
Julia de Le\'on$^{1,2}$.\\
$^{1}$Instituto de Astrof\'{\i}sica de Canarias (IAC), C/V\'{\i}a L\'actea sn, 38205 La Laguna, Spain\\
$^{2}$Departamento de Astrof\'{\i}sica, Universidad de La Laguna, 38206 La Laguna, Tenerife, Spain\\
$^{3}$Astronomical Institute of the Romanian Academy, 5 Cuţitul de Argint, 040557 Bucharest, Romania\\
$^{4}$Faculty of Physics, University of Bucharest, 405 Atomiştilor str., Măgurele 077125, Ilfov, Romania\\
$^{5}$Department of Earth and Planetary Science, The University of Tokyo, Bunkyo, Tokyo, Japan
}

\date{Accepted XXX. Received YYY; in original form ZZZ}

\pubyear{2022}

\begin{document}
\label{firstpage}
\pagerange{\pageref{firstpage}--\pageref{lastpage}}

\maketitle

\begin{abstract}

In the framework of the Visible NEAs Observations Survey (ViNOS) that uses several telescopes at the Canary Islands observatories since 2018, we observed two super fast rotator NEAs,  2021 NY$_1$ and 2022 AB. We obtained photometry and spectrophotometry of both targets and visible spectroscopy of 2022 AB. 
Light curves of 2021 NY$_1$ obtained in 4 different nights between Sept. 30 and Oct. 16, 2021 return a rotation period $P=13.3449\pm0.0013$ minutes and a light curve amplitude $A = 1.00$ mag. We found that 2021 NY$_1$ is a very elongated super fast rotator with an axis ratio $a/b \ge 3.6$. We also report colours $(g-r) = 0.664 \pm 0.013$, $(r-i)  = 0.186 \pm 0.013$, and $(i-z_s) = -0.117 \pm 0.012$ mag. These are compatible with an S-type asteroid.
The light curves of 2022 AB  obtained on Jan. 5 and Jan. 8, 2021 show a rotation period $P=3.0304\pm0.0008$ minutes, with amplitudes $A = 0.52$ and $A =0.54$ mag. 2022 AB is also an elongated object with axis ratio $a/b \ge 1.6$. The obtained colours are $(g-r) = 0.400 \pm 0.017$, $(r-i)  = 0.133 \pm 0.017$, and $(i-z_s) = 0.093 \pm 0.016$. These colours are similar to those of the X-types, but  with an unusually high $(g-r)$ value. Spectra obtained on Jan. 12 and Jan. 14, 2022, are consistent with the reported colours. The spectral upturn over the 0.4 - 0.6 $\mu m $ region of 2022 AB does not fit with any known asteroid taxonomical class or meteorite spectrum, confirming its unusual surface properties.
\end{abstract}




\section{Introduction}

Near-Earth asteroids (NEAs) that graze the Earth at distances of a few Earth-radii are a potential source of concern. They also represent an opportunity to study the smallest Solar System objects from our immediate neighbourhood. Because of their low delta-{\it v} budget they are potential targets for space mission and they can be in-situ resources for space exploration. Thanks to their proximity, their composition, size, rotation properties, and shape can be studied using Earth-based telescopes. These  are obtained by means of spectral and photometric observations during the short window of opportunity when the objects are close enough to Earth.

In 2018, and in the framework of several European funded projects, the Solar System Group of the Instituto de Astrof\'isica de Canarias (IAC), in collaboration with the Astronomical Institute of the Romanian Academy (AIRA), started a joint Visible NEAs Observations Survey (ViNOS). The main goal of the program is to characterize NEAs by using spectroscopic, spectro-photometric, and light curves observations. The program focuses on recently discovered NEAs, those having the smallest sizes, those classified as potentially hazardous (PHAs), possible targets for space missions, and those observed with other techniques (like radar), to provide complementary data. 

We are using the observing facilities at the two observatories in the Canary Islands (Spain) and managed by the IAC, namely the Teide Observatory (OT), located in the island of Tenerife, and the El Roque de los Muchachos Observatory (ORM), in the island of La Palma. One of the key project branch is the use of the world's largest optical telescope, the 10.4 m Gran Telescopio Canarias (GTC), to follow up the most peculiar NEAs.  We also use other 4 to 1 m class telescopes of the ORM. The telescopes we access on a regularly basis from the OT are the 1.5 m Telescopio Carlos S\'anchez (TCS), the 80 cm IAC80 telescope, and two 46 cm robotic telescopes (TAR2 and TAR3). We have already obtained spectroscopic data for more than 100 NEAs, spectrophotometric data for more than 270 NEAs, and light curves of more than 150 NEAs. This is an ongoing survey.

Within ViNOS observed dataset of NEAs we found two objects with extreme/peculiar properties. They were observed on late 2021 and early 2022.  The initial light curves showed they are super fast rotators, with rotation periods of a few minutes. The first one, 2021 NY$_1$, is a Potentially Hazardous Asteroid (PHA) discovered by Pan-STARRS 1 on July 7, 2021. This asteroid has an absolute magnitude $H = 21.84$ that, assuming a geometric albedo $p_V = 0.15$, provides an equivalent diameter of roughly 150 m. The object was at a solar elongation of $\sim 90 \deg$ when it passed through the NASA's WISE space telescope field of view (in August 24, 2021) and it was not detected, which supports that either the asteroid is not significantly larger than 150 m or that it is optically very dark. At the time of our observations 2021 NY$_1$ was on the Goldstone radar target list and it was observed on September 18, 2021. Thus, our observations were also triggered in the framework of the NEOROCKS\footnote{\url{https://www.neorocks.eu/}} project, where we lead the task on the characterization of radar target. When 2021 NY$_1$ was finally observed its rotation period and spectral class were unknown. 

The second object, the Aten-type NEA 2022 AB, was discovered by Piszkesteto Mountain Station observatory on January 2, 2022. This asteroid has an absolute magnitude $H = 23.6$, that suggests a diameter of $\sim 65$ m, assuming a geometric albedo $p_{V} = 0.15$. Asteroid 2022 AB is of interest to NASA because it is a possible target for future missions, so we included it in our target list due to our collaboration with the Near-Earth Object Human Space Flight Accessible Targets Study (NHATS) program\footnote{\url{https://cneos.jpl.nasa.gov/nhats/intro.html}}.

In this paper we show that both NEAs are super fast rotators, with rotation periods of only a few minutes, and we constrain their surface properties. The super fast rotators pose scientific questions because asteroids with a period $< 2.2$ hr (known as the "cohesionless spin-barrier") cannot be merely held together by self-gravity. It is assumed that they must be formed of a contiguous solid or rubble-piles with a significant cohesive strength to resist centrifugal disruption, otherwise they would brake apart \citep{2014M&PS...49..788S,2020MNRAS.495.3990M}. On the other hand, asteroids with periods longer than 2.2 h are typically associated with gravitationally bound aggregates \citep{1996DDA....27..501H}. \citet{2011Icar..214..194H} report that at least two thirds of the asteroids with $H < 20$ have $P < 2.2$ h. 

 The rotation periods give an insight into the body's internal composition and, from its degree of fracture, its collisional history can be inferred. The understanding of the physical properties of these bodies requires multiple observing techniques. 
 
The paper is organized as follows:  Sec. \ref{Sec:Obs} presents the observations and data reduction, Sec. \ref{Sec:phot} describes the methodology to study the light curves and derive information on the asteroid rotation period and shape, and Sec. \ref{Sec:color} discusses the obtained properties from our observations. Finally, Sec. \ref{Sec:conclusions} summarizes the results.

\section{Observations and data reduction}
\label{Sec:Obs}

The time series photometry of NEAs 2021 NY$_1$ and 2022 AB were obtained using four telescopes located at OT, the TCS, the IAC80, and the two TAR2 and TAR3. Low-resolution visible spectroscopy was done using the Intermediate Dispersion Spectrograph (IDS) spectrograph attached to the 2.5~m Isaac Newton Telescope (INT). The observational circumstances are shown in Table \ref{tab_obs_cir}. 


TCS is a 1.52~m telescope with  $f/D =$ 13.8 in a Dall–Kirkham type configuration. It is equipped with the MuSCAT-2 imaging instrument \citep{narita2019}, which is mounted on the Cassegrain focus of the telescope. This instrument allows to obtain simultaneous photometric observations in four broad-band filters, namely \textit{g}~(400 - 550), \textit{r}~(550 - 700), \textit{i}~(700 - 820) and \textit{z$_s$}~(820 - 920) nm. The images were acquired with four independently controllable 1K x 1K CCD  cameras with a pixel scale of 0.435 arcsec/pixel and a FoV of 7.4 x 7.4\,arcmin$^2$ . We used 30 seconds exposure time for each of the four CCDs. The telescope allows only sidereal tracking.

A dedicated pipeline is used to reduce the data from this telescope. As part of the pre-processing, the images are bias and flat-field corrected, and the remaining sky patterns are removed using GNU Astronomy package \citep{gnuastro}. We used the dome flats and biases obtained at the beginning of each observing night using an automated, optimized procedure available from the instrument software control. 

The IAC80 is a 82~cm telescope with  $f/D =$ 11.3 in the Cassegrain focus. It is equipped with the CAMELOT-2 camera, a back-illuminate e2v 4K x 4K pixels CCD of 15 $\mu m^2$ pixels, a plate scale of 0.32 arcsec/pixel, and a field of view of 21.98 x 22.06 arcmin$^2$. We used a Sloan $r$ filter and 60 seconds exposure times with the telescope in sidereal tracking. The images were bias and flat-field corrected in the standard way.

The TAR2 and TAR3 are 46~cm robotic telescopes. TAR2 has a $f/D =$ 2.8 at the prime focus, and is equipped with a FLI-Kepler KL400 camera. It has a back illuminated 2K x 2K pixels CMOS (complementary metal–oxide–semiconductor) with a pixel size of 11 $\mu m^2$. The plate scale is 1.77 arcsec/pixel and the field of view is $\sim$ 1 $\text{deg}^2$. The TAR3 is a twin of TAR2, equipped with a QHY600PRO camera, with a back illuminated 9K x 6K pixels CMOS of 3.76 $\mu m^2$ pixels, a plate scale of 0.65 arcsec/pixel and a field of view of $\sim$ 1.5 x  1.0 $\text{deg}^2$. Both CMOS use a rolling shutter and have the advantage of zero dead-time between images. For a complete description of the QHY600PRO capabilities see \cite{2023arXiv230203700A}.

With the TAR2 we obtained a continuous series of 10 seconds images using a Johnson $V$ filter. Dark and flat-field corrections were applied. The images were astrometrized using {\em astrometry.net} software\footnote{\url{https://astrometry.net/}}, and then each three consecutive images were aligned and combined to produce a final series of images of 30 seconds exposure times using SWARP\footnote{\url{https://www.astromatic.net/software/swarp/}}. In general, the number of images used to obtain the final combined one is determined by the proper motion of the NEA. This is computed such that the total exposure time is shorter than the time it takes for the asteroid trail to be equal to the typical seeing of this telescope ($\sim$ 2 pixels full width at half maximu --FWHM, or 3.6\arcsec). 

In the case of TAR3, we obtained a continuous series of 20 seconds images with no filter. Due to the very low dark current of the QHY camera, we applied only dark and flat-field correction to these images.

Visible spectra of 2022 AB were obtained using the INT. The INT has a 2.54~m diameter primary mirror with a $f/D =$ 15 focal ratio at the Cassegrain focus where the IDS, a long-slit spectrograph, is mounted. IDS was used with the Red+2 detector, a 2K x 4K pixels back-illuminated CCD with 15 $\mu m^2$ pixels which is equivalent to a scale of 0.44 arcsec/pixel. The configuration includes the R150V grating  with the central wavelength 0.65 $\mu$m and a 2\arcsec slit width. This provides high quality spectra over the 0.4-0.9 $\mu$m spectral range with a resolution $R\sim245$ at 0.45 $\mu$m. For this camera the fringing is negligible up to 1 $\mu$m. 

We observed 2022 AB on Jan. 12 and 14, 2022, with the slit oriented at the parallactic angle in order to minimize the effects of atmospheric differential refraction. On Jan. 12 and 14 we obtained 4 spectra of 600 seconds exposure time each night. The strategy was to observe the target as close as possible to the zenith. To get the asteroid's reflectance spectra, we also observed several solar analogue stars, namely HD30246, Hyades 64  \citep{tedesco1982eight} on Jan. 12 and HD70516 and HD98618  \citep{de2014photometric} on Jan. 13, immediately before and after the asteroid with the same configuration and at a similar airmass. The observational circumstances are shown in Table ~\ref{tab_obs_cir}. 

The spectral data reduction was made using the pipeline described by \citet{2019A&A...627A.124P}. It uses GNU Octave software package \citep{octave} to create scripts for IRAF - Image Reduction and Analysis Facility \citep{1986SPIE..627..733T} to perform the tasks automatically. The steps are bias subtraction, flat-field correction, extraction of two dimensional spectral trace from the image to one dimensional spectrum, and the wavelength calibration. The wavelength calibration was done using the emission lines from Cu-Ar and Cu-Ne lamps. The extraction of the raw spectrum from the images was made with the \emph{apall} package. Each image was visually inspected to avoid artifacts. 

\begin{table*}
\caption{Observational circumstances for the asteroids presented in this work. The information includes the object, telescope and filters or grisms used, the date and the starting and end time (UT) of the observations, the phase angle ($\alpha$), and the heliocentric ($r$) and geocentric ($\Delta$) distances of the asteroid at the time of observation.}             
\label{tab_obs_cir}      
\centering          
\begin{tabular}{l c c c c c c c c c}     
\hline     
Asteroid & Telescope & filters/grism & Date & UT (start) & UT (end) & $\alpha$ ($^{\circ}$) & $r$ (au) & $\Delta$ (au) \\ 
\hline                    
   2021 NY$_{1}$ & TAR2 & V & 2021 09 30 & 02:32:47 & 06:15:40 & 66.5  & 1.0173 & 0.0413 &\\
   2021 NY$_{1}$ & TAR2 & V & 2021 10 01 & 02:44:03 & 06:16:54 & 64.4  & 1.0204 & 0.0465 &\\
   2021 NY$_{1}$ & IAC80 & r' & 2021 10 14 & 01:07:04 & 06:06:00 & 46.0 & 1.0737 & 0.1150 &\\
   2021 NY$_{1}$ & TCS & g',r',i',z$_s$ & 2021 10 16 & 04:07:31 & 05:47:06 & 43.6 & 1.0846 & 0.1264 &\\ \hline
   2022 AB & TAR3 & clear & 2022 01 05 & 20:20:15 & 05:13:08 & 22.9 & 1.0344 & 0.0552 &\\
   2022 AB & TCS & g',r',i',z$_s$  & 2022 01 08 & 02:32:47 & 06:15:40 & 16.9 & 1.0296 & 0.0483 &\\
   2022 AB & INT & R150V & 2022 01 12 & 23:14:46 & 23:52:25 & 0.9 & 1.0186 & 0.0351 &\\
   2022 AB & INT & R150V  & 2022 01 14 & 01:11:40 & 03:32:06 & 3.6 & 1.0161 & 0.0326 &\\

\hline                  
\end{tabular}
\end{table*}

\section{Photometry, period analysis and shape constraints}
\label{Sec:phot}

We applied the aperture photometry method to the final images using the Photometry Pipeline\footnote{\url{https://photometrypipeline.readthedocs.io/en/latest/}} (PP) software \citep{2017A&C....18...47M}. PP uses the Source Extractor software for source identification and aperture photometry and the SCAMP\footnote{\url{https://www.astromatic.net/software/scamp/}} software for image registration. Both image registration and photometric calibration are based on matching field stars with the star catalogues (e.g., SDSS, Gaia, URAT-1). Circular aperture photometry is performed using Source Extractor; an optimum aperture radius is identified using a curve-of-growth analysis by the PP. The catalogued stars are used to flux calibrate the results. The images obtained without filter, observed with TAR3 are calibrated to the $r$ SLOAN band using the Pan-STARRS catalogue while the images from the other telescopes are calibrated to the corresponding bands for the filters used. The final calibrated photometry for each field source is written into a queryable database, and target photometric results are extracted from this database. Moving targets are identified using JPL Horizons ephemerides. 

The light curve analysis (including rotation period determination) was carried out using the Tycho package\footnote{\url{https://www.tycho-tracker.com/}}. Magnitudes are (H-G) and light-time corrected and the rotation period is obtained using a Fourier analysis algorithm like in \cite{harris1989}.

\subsection{2021 NY$_1$}

\begin{figure}
	\includegraphics[width=\columnwidth]{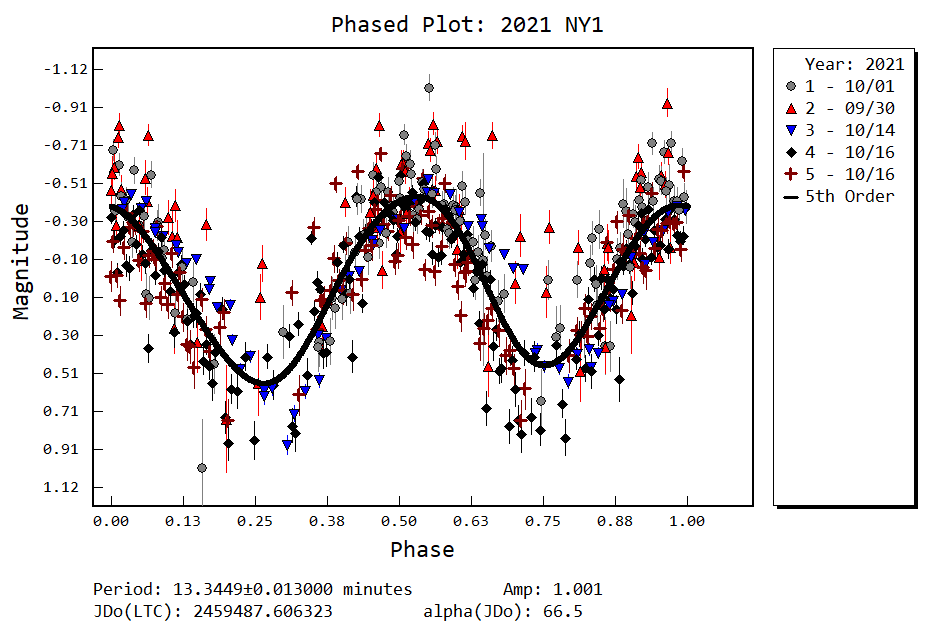}\\
	\includegraphics[width=\columnwidth]{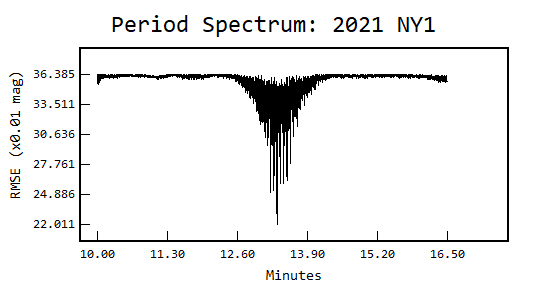}	
    \caption{Upper panel shows the phased composite light curve of 2022 NY$_1$ obtained with our data set. The lower panel is the associated periodogram}
    \label{fig:NY1all}
\end{figure}

The light curves of this asteroid were obtained in four different nights between Sept. 30 to Oct. 16, 2021, using three different telescopes. The analysis of the data using the Tycho package returns a very short rotation period $P=13.3449\pm0.0013$ minutes, and a light curve amplitude $A = 1.00$ mag. The phased light curve and the periodogram are shown Fig. \ref{fig:NY1all}. We used all the data of the four nights, except the data obtained on Oct. 16 with the TCS and the $i$ and $z_s$ filters,  because of its very low signal-to-noise ratio (SNR): the object was barely detected in the individual $i$ and $z_s$ filters 30 seconds images. 

Our result agrees with others' available information. In the framework of the Ondrejov Asteroid Photometry Project, Pravec et al.\footnote{\url{https://www.asu.cas.cz/~ppravec/newres.txt}} observed 2021 NY$_1$ on Sept. 16 and 17 and in Oct. 13, 2021. They report that their Sep. 16 and 17 data give a rotation period of $P=13.3452\pm0.0007$  with an amplitude $A = 0.71$ mag., and including the Oct. 13 data the period is $P=13.34457\pm0.00015$ with $A = 1.01$ mag. Finally, \citet{2021ATel14944....1F} report a period $P=13.3444\pm0.006$ minutes based on photometric data obtained on Sept. 25, 27 and 28, 2021, with a light curve amplitude $A = 1.4$ mag.  

\begin{figure}
	\includegraphics[width=\columnwidth]{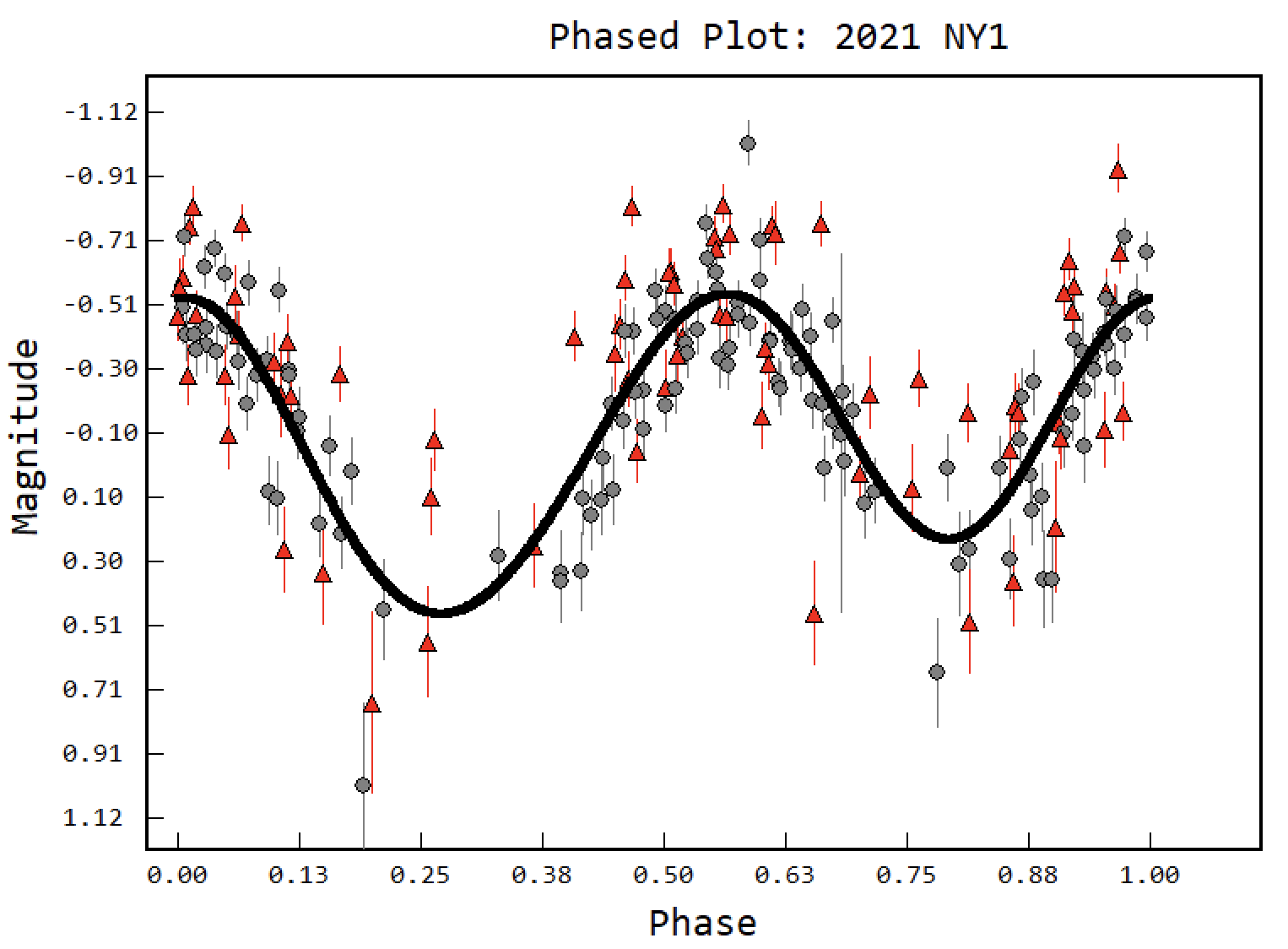}\\
	\includegraphics[width=\columnwidth]{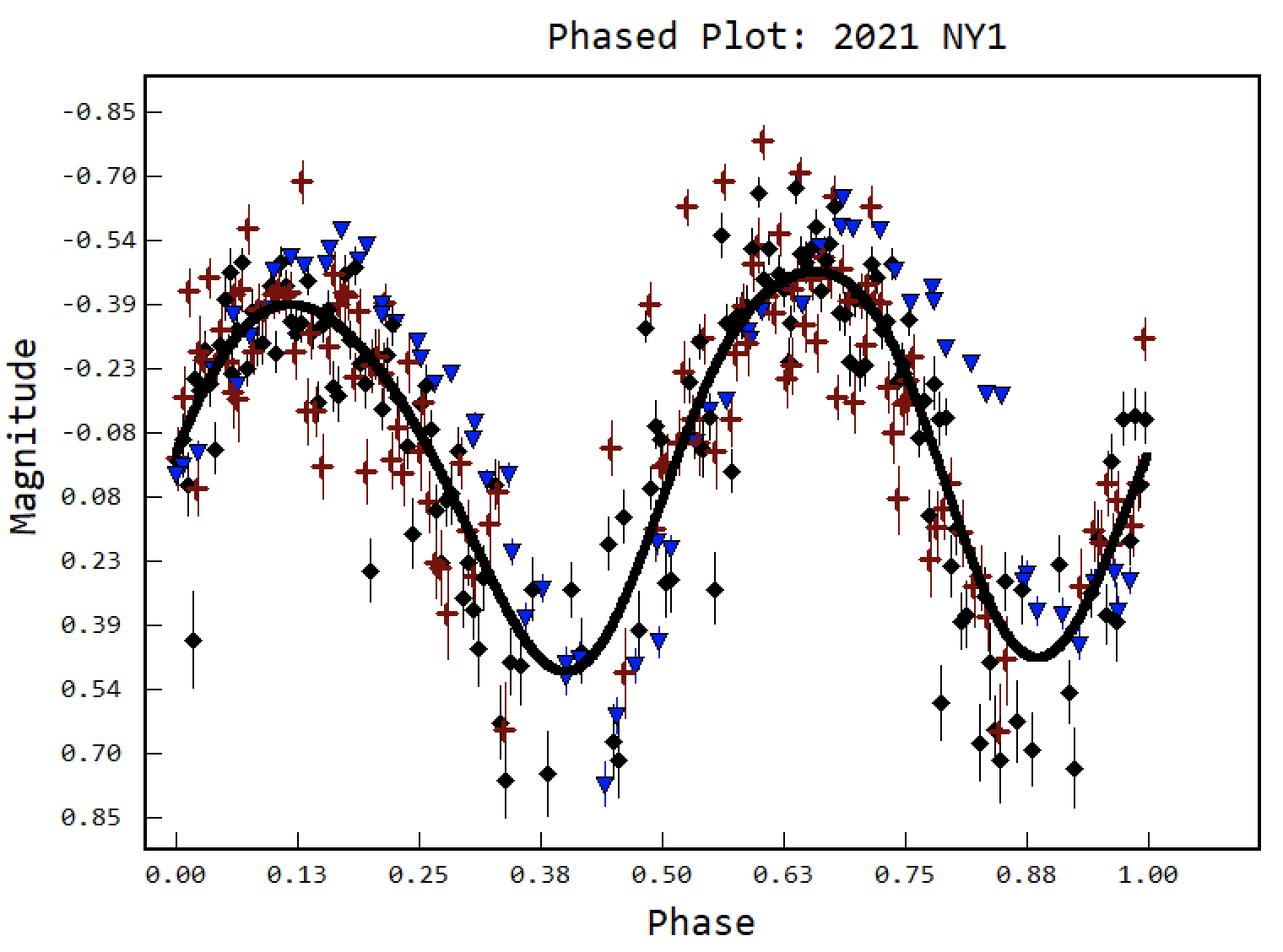}	
    \caption{Upper panel shows the phased composite light curve of 2021 NY$_1$ using TAR2 data only obtained on Sept. 30 and Oct. 1. Lower panel shows the phased composite light curve of 2021 NY$_1$ using IAC80 and TCS data obtained on Oct. 14 and 16, respectively. Data suggest a shape light curve variation from late September to mid October}
    \label{fig:NY1_2epochs}
\end{figure}

While the rotation period is very similar in all the determinations done at different epochs, the reported amplitudes of the light curves are different. They vary from $A=0.71$ on Sep. 16 to $A=1.4$ around Sep. 27 and $A=1.0$ mag. for the data obtained in mid October. Also, the shape of the light curves varied. The phased light curve of the TAR2 nights (Sep. 30 and Oct. 1) seems to present a shape slightly different to that of the IAC80 and TCS ones, obtained in mid October (see Fig. \ref{fig:NY1_2epochs}), with one minima $\sim$0.28 mag. deeper than the other, while the data obtained in mid October present two minima of similar depth. Even if the difference in shape is not large enough to be confirmed within the SNR of the TAR2 data, the light curves obtained by Pravec et al. on Sept. 16-17 and the other on Oct. 13 show a clear change in their shape. This supports our finding, the Sep. 16 light curve shows a large difference between the depth of the two minima. Unfortunately, \citet{2021ATel14944....1F} light curves are not published. The observed light curve shape variations are typically due to changes in the geometry of the observations, in particular, variations of the phase angle ($\alpha$, the Sun-object-Earth angle) and the aspect angle (the angle between the line of sight and the rotation axis of the asteroid). Such changes happen very rapidly during a close encounter like the one of 2021 NY$_1$ on Sep. 22, 2021. The amplitude vs. phase relation was described in \cite{1990A&A...231..548Z} where they show that it is almost linear for 0$^{\circ}$ $< \alpha <$ 60$^{\circ}$, but then it behaves in a much more complex way. The data presented in this paper was obtained with large $\alpha$ values (between 44$^{\circ}$ and 67$^{\circ}$), but Pravec's data obtained in mid September was obtained at $\alpha$ angles between 111$^{\circ}$ and 113$^{\circ}$ and \citet{2021ATel14944....1F} data at $\alpha$ angles between 71$^{\circ}$ and 86$^{\circ}$. Therefore, the interpretation of this data is not straightforward.

A further analysis of all the complete set of light curves using, for example, light curve inversion techniques \citep{2001Icar..153...24K,2001Icar..153...37K}, and its combination with radar data modelling, can provide more information on the shape of the asteroid and its pole axis orientation. In any case, the amplitude determined from our data can be used to put some constraints on its shape. Assuming that the asteroid is a triaxial ellipsoid with axes $a, b, c$, we can calculate a limit for the axis ratio from $a/b \ge 10^{0.4A}$. According to our amplitude determination, $a/b \ge 2.5$. Using the amplitude determination from \citet{2021ATel14944....1F} the ratio is even larger ($a/b \ge 3.6$). We conclude that 2021 NY$_1$ is a very elongated super fast rotator.

\subsection{2022 AB}
\begin{figure}
	\includegraphics[width=\columnwidth]{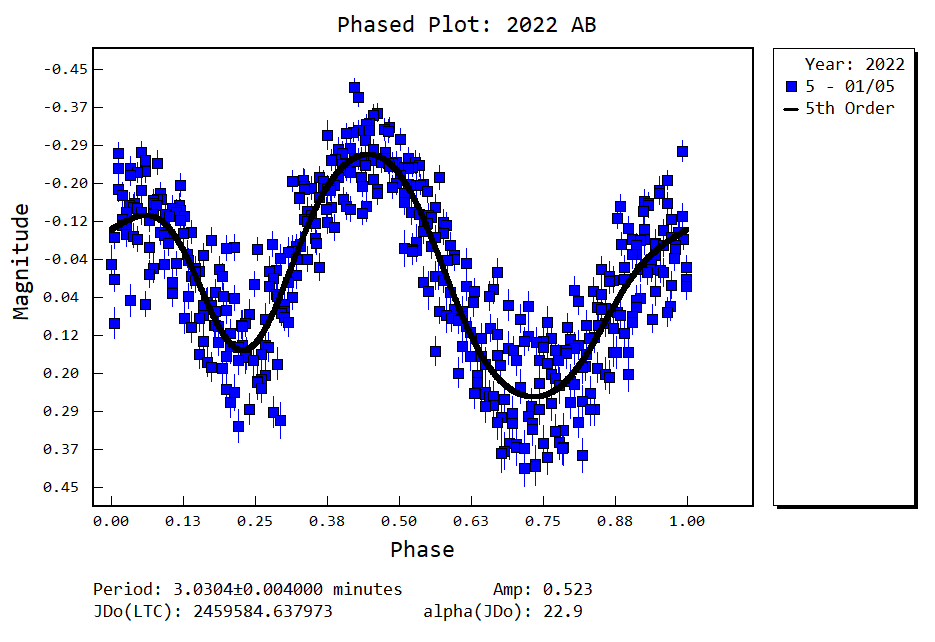}\\
	\includegraphics[width=\columnwidth]{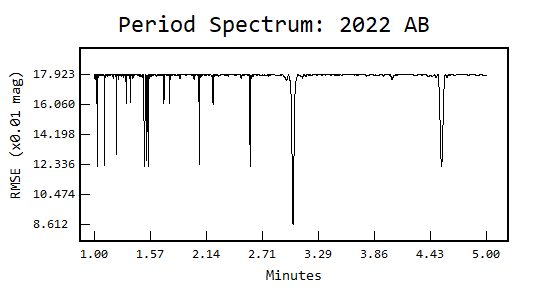}	
    \caption{Upper panel shows the phased composite light curve obtained using TAR3 data from Jan.5, 2022. Lower panel is the associated periodogram computed using Tycho.}
    \label{fig:AB_TAR3all}
\end{figure}

\begin{figure}
	\includegraphics[width=\columnwidth]{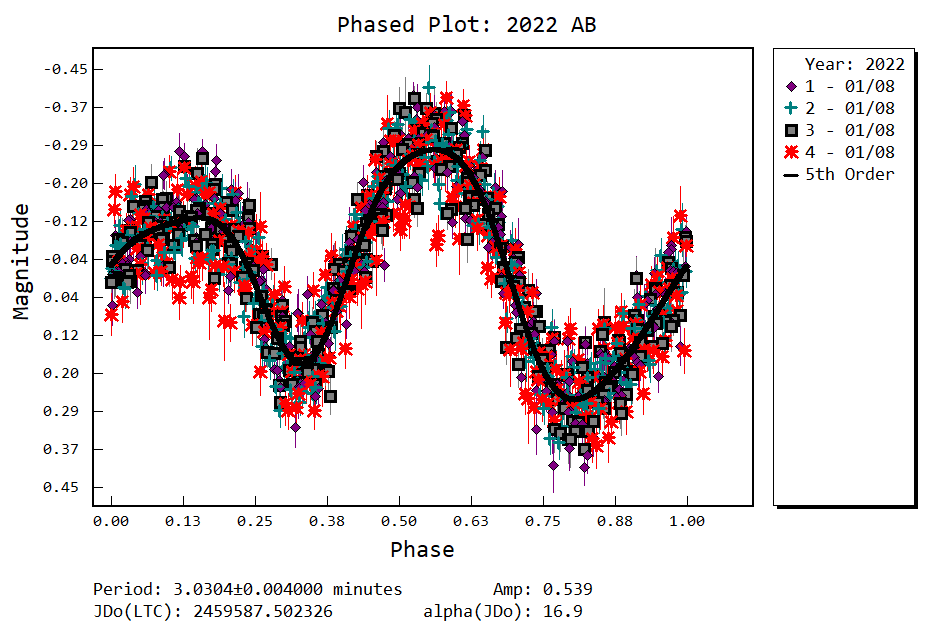}\\
	\includegraphics[width=\columnwidth]{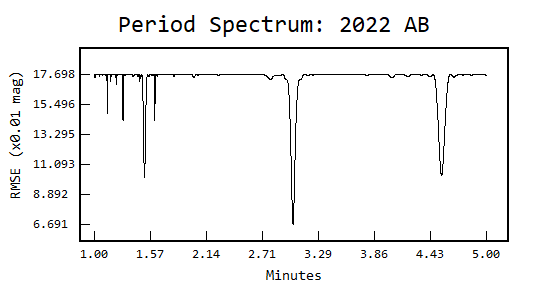}	
    \caption{Upper panel shows the phased composite light curve obtained using TCS data from Jan. 8, 2022. Magnitudes are distance and phase corrected and all the filters are colour corrected using the colours. Lower panel is the associated periodogram computed using Tycho.}
    \label{fig:AB_TCSall}
\end{figure}

The light curves of this asteroid were obtained in two different nights, Jan. 5 and 8, 2022, using TAR3 and TCS telescopes. The analysis of the data done separately for each observing night (see periodograms and phased light curves in Fig. \ref{fig:AB_TAR3all} and Fig. \ref{fig:AB_TCSall}) returns a very short rotation period $P=3.0304\pm0.0008$ minutes, exactly the same value on both nights. 

Pravec et al. and \cite{2022EPSC...16.1161K} also report time-series photometric observations of this asteroid. Pravec et al. observed it during five nights, Jan. 7, 8, 9, 10 and 11, 2022 and they report a period $P=3.03088\pm0.00006$ minutes. \cite{2022EPSC...16.1161K} observed 2022 AB from 4 Jan to 26 Jan, and report a preliminary determination of $P=3.033\pm0.002$ minutes based on the light curves obtained on Jan. 11, 2022 only.

The amplitudes of the light curves measured in both nights are almost the same, $A = 0.52$ and $A =0.54$ magnitudes in Jan. 5 and 8 respectively. Pravec et al. and \cite{2022EPSC...16.1161K} report a very similar amplitude ($A = 0.55$ and $0.51 < A < 0.55$ magnitudes respectively). This suggests that the aspect angle did not varied much between Jan. 5 and Jan. 11, a few days before the asteroid flown by close to Earth on Jan. 20, 2022. The derived axis ratio of 2022 AB is $a/b \ge 1.6$. 

\subsection{Super fast rotators}

\begin{figure}
\begin{center}
\includegraphics[width=8cm]{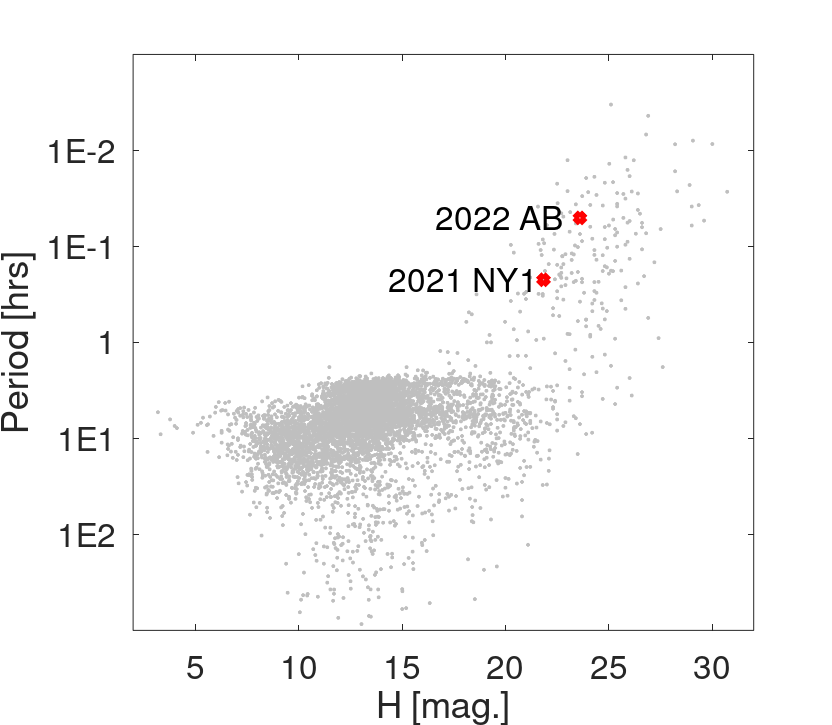}
\end{center}
\caption{Distribution of rotation period versus absolute magnitude ($H$) of 5569 asteroids using the data retrieved from LCDB. For comparison, the two objects studied in this work are shown with red symbols.}
\label{SpinvsFrequency}
\end{figure}

In order to compare the rotational properties of these two NEAs with those available in the literature we used the Asteroid light curve Database\footnote{\url{https://minplanobs.org/mpinfo/php/lcdb.php}} LCDB \citep[][accessed on July 10, 2022]{2009Icar..202..134W}. We selected the asteroids flagged with $U~\geq~2-$ (the threshold recommended by the authors for statistical studies), were $U$ represents the assessment of the quality of the period solution reported by \citep{2009Icar..202..134W}. The results are shown in Fig.~\ref{SpinvsFrequency}.

Most small bodies of our Solar System have rotation periods between 2.2 and 20 hours. As of 2022, a group of $\sim$ 1\,600 asteroids out of 32\,249 with accurate period determination have an estimated rotation period below 2.2 hours, and only 261 have rotation periods smaller than 13.5 minutes, like the NEAs presented in this paper.  Looking at the literature \citep[][and references therein]{2005MNRAS.359.1575L, 2010A&A...517A..23D, 2011A&A...535A..15P, 2014A&A...572A.106P, 2018P&SS..157...82P, 2019Icar..324...41B, 2019AJ....158..196D, 2019A&A...627A.124P, 2021MNRAS.508.1128S, 2022A&A...665A..26M} we found that 53 asteroids of that sample have taxonomical classes determined, and 41 of them are stony NEAs (S- or Q-type), all with absolute magnitudes $H>20$ (with equivalent diameter $< 340$ m assuming $p_V = 0.15$). This strong bias against rapid rotation was taken as evidence that the large majority of asteroids were strengthless "rubble piles"  (conglomerations of smaller pieces, loosely coalesced under the influence of gravity) that did not possess any inherent tensile strength \citet{1996DDA....27..501H}. 
Rotation periods significantly below the cohesion-less spin-barrier are indicative of intrinsic strength in the asteroids \citep{2000Icar..148...12P}, that means they are coherent bodies or monoliths, and predominate in the sub-km size population of NEAs. \citet{2011Icar..214..194H} report that at least two thirds of the asteroids with $H < 20$ have $P < 2.2$ h. On the other hand, \cite{2014M&PS...49..788S} show  that "the finest grains
within an asteroid can serve as a cement, a cohesive matrix that binds larger boulders together into a body, allowing it to spin more rapidly than the surface disruption limit". They conclude that the super fast rotators population could consist of monolithic bodies, as well as rubble-pile asteroids with a significant cohesive strength to resist centrifugal disruption.

The rotation of asteroids has been set and altered by several processes during their formation and evolution, e.g. collisions and micro collisions and YORP effect \citep{2000Icar..148....2R}. Sub-km asteroids are the small pieces produced by the successive collisions of larger objects and, due to their small size, their rotation is largely affected by YORP effet. The study of their physical properties is of fundamental importance to understand all these processes.  

\section{colours, spectrum \& taxonomical classification}
\label{Sec:color}

Simultaneous calibrated images in $g$, $r$, $i$, and $z_s$ obtained with the TCS are used also to obtain the colours of both asteroids and to determine its taxonomic classifications. For 2022 AB the accurate colours were retrieved by computing the offsets between the phased light curves obtained with the TCS (see Table \ref{TCSobslog}). 

In the case of 2021 NY$_1$ the accurate colours couldn't be retrieved from the individual images because of the poor SNR of the images in the $i$ and $z_s$ filters. To compute colours we combined images in order to have better SNR by using  the following procedure: we first registered all the individual images using the Photometry Pipeline; we obtained co-added images by combining  9 consecutive images (9 x 30 s = 270 s equivalent exposure time) in the object moving frame (the object is point source, the stars are trails) and we measured the brightness of the object on each of these images using aperture photometry. Then, we co-added every 9 consecutive images in the sky frame (object is trail, stars are point like sources) and using Photometry Pipeline we derived the Zero Points corresponding to the same apertures used on the asteroid combined images. Finally, we averaged all the so determined magnitudes in the four filters (about 20 data points), removed some outliers and derived the colours presented in Table \ref{TCSobslog}.

\begin{table*}
	\caption{TCS/MuSCAT2 colours and taxonomical classification. Table includes the asteroid designation, the observation duration in hours ($t_{obs}$), the total number of images acquired for one channel ($N_i$), the median colours and their equivalent errors, and the resulting taxonomic classification using the K-Nearest neighbours (Tax$_{KNN}$) and Random Forest (Tax$_{RF}$) algorithms. KNN$_{prob}$ is the clarification probability. Finally, the last column contains the final attributed spectral type.}
    \begin{tabular}{l c c c c c c c c c c}
    \hline    
Asteroid &  $t_{obs}$ (h) & $N_{i}$ & $(g-r)$ &  $(r-i)$ & $(i-z_s)$ & Tax$_{KNN}$ & KNN$_{prob}$ & Tax$_{RF}$ & RF$_{prob}$ & Taxonomy\\    \hline   
    2021 NY$_1$ &  1.6511 & 153 & 0.664 $\pm$ 0.013 & 0.186 $\pm$ 0.013 & -0.117 $\pm$ 0.012 & S & 0.915 & S & 0.992 & S \\
    2022 AB  &  1.3561 & 304 & 0.400 $\pm$ 0.017 & 0.133 $\pm$ 0.017 & 0.093 $ \pm$ 0.016 & C & 0.819 & X & 0.920 & X \\
	\end{tabular}
	\label{TCSobslog}
\end{table*}

To perform the taxonomic classification of 2021 NY$_1$ and 2022 AB, we used the method and results in Popescu et al. (in preparation). We  used both, the K-Nearest neighbours (KNN) and Random Forest (RF) algorithms. Each algorithm was implemented using the {\tiny SCIKIT-LEARN} package from {\tiny PYTHON}. The KNN algorithm classifies an object based of the label values or taxonomy of its neighbours in the colour-colour space, while the Random Forest assigns the final label of an object using decision-tree structures, which are also drawn using objects with known taxonomy.

Thus, the first step was to built a training set, namely a set of objects for which we know both the photometric colours (from our data set) and spectral data (from the literature). We searched the available spectral information of all the objects with TCS colours in different public databases, from SMASS-MIT-Hawaii Near-Earth Object Spectroscopic Survey (MITHNEOS MIT-Hawaii Near-Earth Object Spectroscopic Survey) program \citep{2019Icar..324...41B} and from Modeling for Asteroids (M4AST) database \citep{2012A&A...544A.130P} and identified spectral data for 84 of the objects we observed with TCS. In order to increase the training sample we computed the synthetic colours  using  the visible spectra published by \citet{2019A&A...627A.124P} and \citet{2018P&SS..157...82P}. Thus, the final training sample consists of 154 asteroids divided as 5 A-type, 8 V-type , 34 Q-types and 48 S-complex, 7 B-types, 15 C-complex, 9 D-types, and 28 X-complex.

\begin{figure}
	\centering 
    \includegraphics[width = 8.87cm]{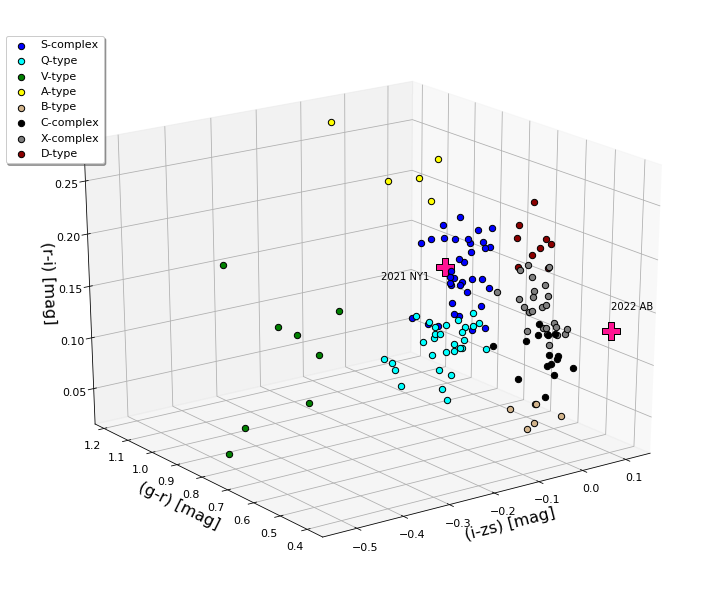}
	\caption{The $(g-r)$ vs. $(r-i)$ vs. $(i-z_s)$ 3D-colour diagram of all objects with known spectral classification, used as training data for the two pattern recognition algorithms used for classification. The taxonomic types defined in \citet{2009Icar..202..160D} system have been divided in three major compositional groups, namely the Q\,/\,S-complex (green and blue dots), C-complex (black dots) and X-complex (grey dots). Besides them, three end-member types are considered, A-, D- and V-type. Both, 2021 NY1 and 2022 AB are drawn as purple crosses.}
	\label {rmi_vs_imz}
\end{figure}

 Since the number of features on which the spectral classification has been based is low, we chose to implement the KNN algorithm so that the taxonomy is given by calculating the euclidean distance to the first three nearest neighbours. Also, because the assigned spectral type is sensitive to the position of each object relative to the training set, we needed to account for the magnitude errors. To do that, we started from the colour value and its error and generated three normal distributions (one for each colour) of 10\,000 fictitious colour values. Then, in each of these cases we classified the object. Finally, the assigned taxonomy was the one with the highest frequency.

For the classification based on the Random Forest algorithm, we used nine different decision-trees, each of them with a maximum of 50 leaf nodes. To account for the magnitude errors, a procedure similar to the one for the KNN method has been used. We applied the RF algorithm for each of the 10\,000 colour sets and chose the predicted taxonomy with the highest frequency.

In the end, having the output of each algorithm, we needed to pick a final taxonomy. For that, we compared the prediction probability of each method and chose the taxonomy corresponding to the algorithm having a higher probability. We emphasize that the goal of these classification methods was to constrain the compositional group of these objects, namely to distinguish between carbonaceous, silicate or basaltic composition.  We found that the colours of 2021 NY$_1$ are fully consistent with an S-type classification (see Fig.\ref{rmi_vs_imz}) so we can discard it as optically very dark. So, assuming a geometric albedo $p_V = 0.23$, the mean albedo of the S-complex according to \cite{2011ApJ...741...90M}, and using the reported absolute magnitude $H = 21.84$, its diameter should be $D < 120$m. On the other hand, 2022 AB in Fig.\ref{rmi_vs_imz} is at the border of C- and X-complexes, with a higher probability for being an X-type object.  Nevertheless, for this object we notice the small value of $(g-r) = 0.400\pm0.017$ which suggests an unusual spectral upturn in the 0.4 - 0.55 $\mu m$ region. 

\cite{2022EPSC...16.1161K} also reported photometric colours for 2022 AB obtained from 6 different sets of observations. The mean value of the colours obtained with this data is $(g-r) = 0.35 \pm 0.03$, $(r-i)  = 0.13 \pm 0.02$, and $(i-z_s) = 0.08 \pm 0.02$. These colours are almost identical to those reported in this paper within the uncertainties. \cite{2022EPSC...16.1161K} transformed colours into reflectances and suggested that 2022 AB is a Cb-type asteroid because they claim it presents the characteristic visible absorption around 0.6 $\mu m$.    

\begin{figure}
	\includegraphics[width=\columnwidth]{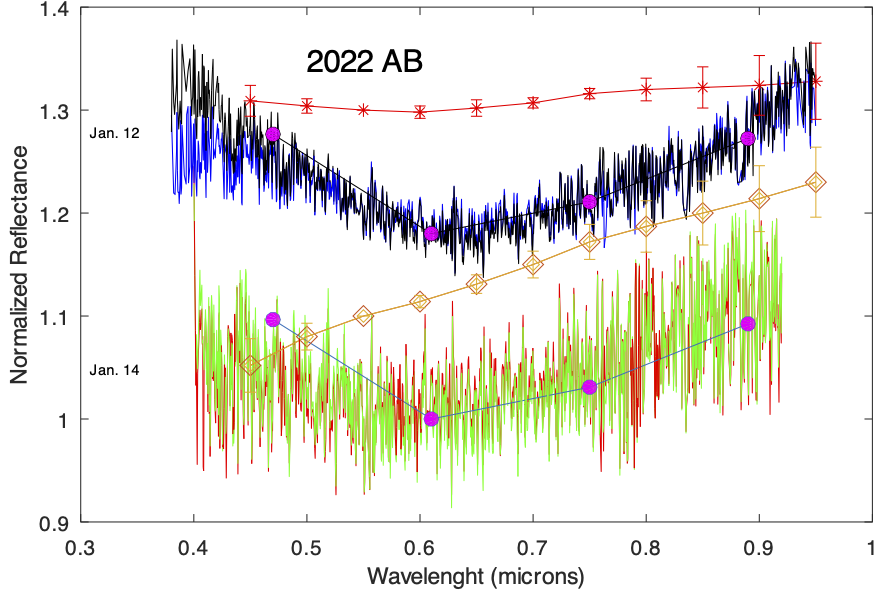}	
   \caption{Spectra of 2022 AB obtained on Jan. 12 and Jan. 14 nights. In black and blue, shifted 0.18 in the vertical axis for clarity, are the reflectance spectra obtained on Jan. 12, in black the reflectance using Hyades 64 and in blue using HD30246 solar analogue stars. In green and red are the reflectance spectra obtained on Jan. 14. The red curve represents the reflectance spectrum using HD98618 and in green using HD70516. Over-plotted, in pink filled circles, are the reflectances obtained using the colours measured with the TCS. In brown, represented by diamonds and shifted in y-axis by 0.1 for clarity, is the template spectrum of the X-type taxonomy and in red, represented by stars and shifted in y-axis by 0.3, is the template specutrm of the Cb-type from the \citet{2009Icar..202..160D} taxonomy. Notice that the spectrum of 2022 AB doesn't fit with X- or Cb- types.}
   \label {spectra}
\end{figure}

In the case of asteroid 2022 AB we also obtained low-resolution visible spectra during two nights, Jan. 12 and Jan. 14, 2022 using the INT. The observations, image reduction procedure and spectra extraction are described in Sec. \ref{Sec:Obs}. The spectral reflectance was obtained by: 1) averaging the spectra of each object (asteroid and solar analogues) to obtain the final extracted spectrum; 2) dividing the averaged extracted asteroid spectrum by the averaged spectrum of each solar analogue star observed the same night; 3) normalizing the result to unity at 0.55 $\mu$m. In Fig. \ref{spectra} we show the so obtained reflectance spectra of 2022 AB. 

In order to compare the spectral data with the spectro-photometric observations we converted the colours obtained with the TCS in reflectances and plotted them in Fig. \ref{spectra}. This was done using the Eq.~\ref{color2reflec} as we did e.g. in \citet{2018A&A...617A..12P}. 

\begin{equation}
{R_{aster}^{f1}}/{R_{aster}^{f2}} = 10^{-0.4(C_{f1-f2} - C_{f1-f2}^{Sun})}
\label{color2reflec}
.\end{equation}
were $f1$ and $f2$ are two different filters, the corresponding colour is $C_{f1-f2}$, $C_{f1-f2}^{Sun}$ represents the colour of the Sun, and ${R_{aster}^{f1}}$ and ${R_{aster}^{f2}}$ are the asteroid reflectances. The following colours of the Sun were used $(g-r)^{Sun}$ = 0.50, $(r-i)^{Sun}$ = 0.10, $(i-z_s)^{Sun}$ = 0.03. These values (Popescu et al., in preparation) were derived for the filters available for MuSCAT2 instrument and are comparable to those provided by \citet{2006MNRAS.367..449H} for the SDSS filters $(g - r)^{Sun}$ = 0.45 $\pm$ 0.02, $(r - i)^{Sun}$ = 0.12 $\pm$ 0.01, $(i - z)^{Sun}$ = 0.04 $\pm$ 0.02. Finally, we normalized the reflectances at the $r$ filter (${R_{aster}^{r}}$ = 1). The results are shown in Fig. \ref{spectra} as pink filled circles, overplotted to the obtained reflectance spectra.

The four spectra are very similar, presenting a similar curved shape with a minimum at $\sim 0.6 \mu$m. At wavelengths larger than 0.6 $\mu$m the four spectra are almost equal, small differences appear only in the near-UV region (0.4-0.6 $\mu$m) being the most relevant difference the near-UV slope. Spectra obtained on Jan. 12 are slightly bluer than those obtained on Jan. 14. Note that the reflectance spectra of asteroids in the near-UV region are very sensitive to the used solar analogue star, as we shown in \citet{2022A&A...664A.107T}. Following the criterion used in Tatsumi et al., for the analysis, we use the spectra obtained using Hyades 64 as the most reliable one. The first important conclusion is that the colours agree with the obtained spectra. The spectral curve clearly shows the upturn in the 0.4 - 0.6 $\mu m$ region confirming that the reflectance properties of 2022 AB are unusual. 

By using various curve matching methods, we found that 2022 AB spectrum does not fit with any known asteroid taxonomical class or meteorite spectrum, not even with the Cb class as suggested by \cite{2022EPSC...16.1161K}. To determine the taxonomic class we used the M4AST\footnote{\url{http://m4ast.imcce.fr/}} on-line tool \citep{2012A&A...544A.130P}. The classification was performed in the framework of the Bus-DeMeo taxonomy  \citep{2009Icar..202..160D}. None of the Bus-DeMeo spectral classes fits with the whole spectrum of 2022 AB. Only if we limit the wavelength range to the 0.6-0.9 $\mu$m region we obtain a result that is in agreement with the assigned taxonomy from the colours: it fits to an X-type asteroid. The X-type taxonomical class was introduced by the Bus taxonomy, and it groups together three Tholen classes \citep{1984PhDT.........3T}, i.e., P-, M-, and E-types, that can only be separated using the albedo. Therefore, having the same spectral shape in the visible wavelength region, X-types include very dark and primitive objects, like the P-types, and moderately to very bright asteroids like M-types or E-types. In the event of 2022 AB being an X-type with low albedo, the hypothesis of being a rubble-pile asteroid with a significant cohesive strength to resist centrifugal disruption \citep{2014M&PS...49..788S} will be supported. If super fast rotation implies instead a monolithic asteroid as suggested by \cite{2000Icar..148...12P}, then 2022 AB is less likely to be a dark X-type asteroid. Alternatively, M-type will be a good candidate, as they are presumed to be the progenitors both of differentiated iron-nickel meteorites and enstatite chondrites \citep{1976JGR....81..905G,1990JGR....95.8323C,1993Icar..106..573G}. These M-type asteroids present a high bulk density (e.g. 16 Psyche density is $7.6\pm3.0$ g/cm$^3$ according to \citealt{2008Icar..195..184S}). In any case, none of them presents the pronounced curvature shown in the spectrum of 2022 AB.

Trying to obtain information on the surface composition of 2022 AB we also compared the spectrum with laboratory spectra of meteorites using M4AST interface. M4AST operates with more than 2\,500 meteorites spectra provided in Relab database\footnote{\url{http://www.planetary.brown.edu/relab/}} \citep{2004LPI....35.1720P, 2016LPI....47.2058M}. We didn't found any meteorite with a similar spectrum.  

In the case of 2022 AB, the impossibility to obtain a taxonomical classification implies that we cannot constrain its geometric albedo and thus make a reliable estimation of its size. Unfortunately, further analysis and more data (spectra in other wavelength ranges, albedo, etc) is needed to know the composition of 2022 AB and thus obtain information about its internal structure, but this object will not be well suited for Earth-based observations for decades.


\section{Conclusions}
\label{Sec:conclusions}

We present photometry and spectrophotometry of two super fast rotator NEAs,  2021 NY$_1$ and 2022 AB. The data were obtained using four telescopes (1.5m TCS, 0.8m IAC80, 0.46m TAR2 and 0.46m TAR3) located at Teide Observatory (Tenerife, Canary Islands, Spain). In the case of 2022 AB, we also present low resolution spectroscopy in visible wavelengths observed with the 2.5m INT telescope at El Roque de los Muchachos Observatory (La Palma, Canary Islands, Spain). 

The light curves of 2021 NY$_1$ obtained in four different nights between Sept. 30 and Oct. 16, 2021 return a very short rotation period $P=13.3449\pm0.0013$ minutes and a light curve amplitude $A = 1.00$ mag. This result is consistent with those from Pravec et al. ($P=13.3452\pm0.0007$ and $P=13.34457\pm0.00015$ minutes) and \citet{2021ATel14944....1F} ($P=13.3444\pm0.006$ minutes). From the reported amplitudes, even if they can be slightly overestimated because the object was observed at very large phase angles, we conclude that 2021 NY$_1$ is a very elongated super fast rotator with an axis ratio $a/b \ge 3.6$. From the observations obtained simultaneously with $g,~r,~i,~z_s$ filters made with the MuSCAT2/TCS instrument on Oct. 16, 2021 we report 2021 NY$_1$ colours $(g-r) = 0.664 \pm 0.013$, $(r-i) = 0.186 \pm 0.013$, and $(i-z_s) = -0.117 \pm 0.012$. These values are compatible with an S-type asteroid. Assuming a geometric albedo $p_V = 0.23$ and its reported absolute magnitude $H = 21.84$, 2021 NY$_1$ diameter should be $D < 120$m.

The light curves of 2022 AB obtained in two different nights, Jan. 5 and Jan. 8, 2021 return a super rotation period $P=3.0304\pm0.0008$ minutes, with amplitudes $A = 0.524$ and $A  = 0.54$. Similar results are reported by Pravec et al. and \cite{2022EPSC...16.1161K}. 2022 AB is also an elongated object with its axis ratio $a/b \ge 1.6$. The obtained colours using the TCS are $(g-r) = 0.400 \pm 0.017$, $(r-i) = 0.133 \pm 0.017$, and $(i-z_s) = 0.093 \pm 0.016$. The closer taxonomic class using the obtained colours corresponds to an X-type asteroid, but with an unusually high $(g-r)$ value. Spectral data were obtained during two nights, Jan. 12 and Jan. 14, 2022 using the INT/IDS spectrograph. The asteroid spectrum presents an unusual upturn in the blue wavelengths, but it is consistent with the colours obtained with TCS. The reflectance curve does not fit with any known asteroid taxonomic class or meteorite spectrum, which makes it very difficult to draw conclusions about the composition (and diameter) of this asteroid.

Finally, for 2021 NY$_1$ we determined an upper limit of its equivalent diameter, $D < 120$m,  assuming a geometric albedo $p_V = 0.23$ (the mean albedo of the S-complex) and an absolute magnitude $H = 21.84$. For 2022 AB the situation is much more complex because it is impossible to obtain a reliable taxonomical classification which implies that we have no idea of its geometric albedo.

\section*{Acknowledgements}

This project has received funding from the European Union's Horizon 2020 research and innovation program under grant agreement No 870403 (NEOROCKS). JL, MP, and JdL acknowledge support from the ACIISI, Consejer\'{\i}a de Econom\'{\i}a, Conocimiento y Empleo del Gobierno de Canarias and the European Regional Development Fund (ERDF) under grant with reference ProID2021010134. DM, FTR, JL, and JdL acknowledge support from the Agencia Estatal de Investigacion del Ministerio de Ciencia e I\'nnovacion (AEI-MCINN) under grant "Hydrated Minerals and Organic Compounds in Primitive Asteroids" with reference PID2020-120464GB-100. The work of MP was supported by a grant of the Romanian National Authority for  Scientific  Research -- UEFISCDI, project number PN-III-P1-1.1-TE-2019-1504. 

This article is based on observations made with the Telescopio Carlos S\'anchez, IAC80, TAR2 and TAR3 telescopes operated on the island of Tenerife by the Instituto de Astrof\'{\i}sica de Canarias in the Spanish Observatorio del Teide and the Isaac Newton Telescope operated by the Isaac Newton Group on the island of La Palma in the Spanish Observatorio del Roque de los Muchachos.

We finally thanks Filipe Monteiro for his very useful comments.



\section*{Data Availability}

The data underlying this article will be shared on reasonable request to the corresponding author.



\bibliographystyle{mnras}
\bibliography{Licandro_FastRotators_rev1} 

\begin{thebibliography}{}
\makeatletter
\relax
\def\mn@urlcharsother{\let\do\@makeother \do\$\do\&\do\#\do\^\do\_\do\%\do\~}
\def\mn@doi{\begingroup\mn@urlcharsother \@ifnextchar [ {\mn@doi@}
  {\mn@doi@[]}}
\def\mn@doi@[#1]#2{\def\@tempa{#1}\ifx\@tempa\@empty \href
  {http://dx.doi.org/#2} {doi:#2}\else \href {http://dx.doi.org/#2} {#1}\fi
  \endgroup}
\def\mn@eprint#1#2{\mn@eprint@#1:#2::\@nil}
\def\mn@eprint@arXiv#1{\href {http://arxiv.org/abs/#1} {{\tt arXiv:#1}}}
\def\mn@eprint@dblp#1{\href {http://dblp.uni-trier.de/rec/bibtex/#1.xml}
  {dblp:#1}}
\def\mn@eprint@#1:#2:#3:#4\@nil{\def\@tempa {#1}\def\@tempb {#2}\def\@tempc
  {#3}\ifx \@tempc \@empty \let \@tempc \@tempb \let \@tempb \@tempa \fi \ifx
  \@tempb \@empty \def\@tempb {arXiv}\fi \@ifundefined
  {mn@eprint@\@tempb}{\@tempb:\@tempc}{\expandafter \expandafter \csname
  mn@eprint@\@tempb\endcsname \expandafter{\@tempc}}}

\bibitem[\protect\citeauthoryear{{Akhlaghi} \& {Ichikawa}}{{Akhlaghi} \&
  {Ichikawa}}{2015}]{gnuastro}
{Akhlaghi} M.,  {Ichikawa} T.,  2015, \mn@doi [ApJS]
  {10.1088/0067-0049/220/1/1}, \href
  {https://ui.adsabs.harvard.edu/abs/2015ApJS..220....1A} {220, 1}

\bibitem[\protect\citeauthoryear{{Alarcon}, {Licandro}, {Serra-Ricart},
  {Joven}, {Gaitan}  \& {de Sousa}}{{Alarcon}
  et~al.}{2023}]{2023arXiv230203700A}
{Alarcon} M.~R.,  {Licandro} J.,  {Serra-Ricart} M.,  {Joven} E.,  {Gaitan} V.,
    {de Sousa} R.,  2023, \mn@doi [arXiv e-prints] {10.48550/arXiv.2302.03700},
  \href {https://ui.adsabs.harvard.edu/abs/2023arXiv230203700A} {p.
  arXiv:2302.03700}

\bibitem[\protect\citeauthoryear{{Binzel} et~al.,}{{Binzel}
  et~al.}{2019}]{2019Icar..324...41B}
{Binzel} R.~P.,  et~al., 2019, \mn@doi [\icarus]
  {10.1016/j.icarus.2018.12.035}, \href
  {https://ui.adsabs.harvard.edu/abs/2019Icar..324...41B} {324, 41}

\bibitem[\protect\citeauthoryear{{Cloutis}, {Gaffey}, {Smith}  \&
  {Lambert}}{{Cloutis} et~al.}{1990}]{1990JGR....95.8323C}
{Cloutis} E.~A.,  {Gaffey} M.~J.,  {Smith} D.~G.~W.,   {Lambert} R. S.~J.,
  1990, \mn@doi [\jgr] {10.1029/JB095iB06p08323}, \href
  {https://ui.adsabs.harvard.edu/abs/1990JGR....95.8323C} {95, 8323}

\bibitem[\protect\citeauthoryear{{DeMeo}, {Binzel}, {Slivan}  \& {Bus}}{{DeMeo}
  et~al.}{2009}]{2009Icar..202..160D}
{DeMeo} F.~E.,  {Binzel} R.~P.,  {Slivan} S.~M.,   {Bus} S.~J.,  2009, \mn@doi
  [\icarus] {10.1016/j.icarus.2009.02.005}, \href
  {https://ui.adsabs.harvard.edu/abs/2009Icar..202..160D} {202, 160}

\bibitem[\protect\citeauthoryear{{Devog{\`e}le} et~al.,}{{Devog{\`e}le}
  et~al.}{2019}]{2019AJ....158..196D}
{Devog{\`e}le} M.,  et~al., 2019, \mn@doi [\aj] {10.3847/1538-3881/ab43dd},
  \href {https://ui.adsabs.harvard.edu/abs/2019AJ....158..196D} {158, 196}

\bibitem[\protect\citeauthoryear{Eaton, Bateman, Hauberg  \& Wehbring}{Eaton
  et~al.}{2020}]{octave}
Eaton J.~W.,  Bateman D.,  Hauberg S.,   Wehbring R.,  2020, {GNU Octave}
  version 6.1.0 manual: a high-level interactive language for numerical
  computations.
\url {https://www.gnu.org/software/octave/doc/v6.1.0/}

\bibitem[\protect\citeauthoryear{{Ferrais} \& {Jehin}}{{Ferrais} \&
  {Jehin}}{2021}]{2021ATel14944....1F}
{Ferrais} M.,  {Jehin} E.,  2021, The Astronomer's Telegram, \href
  {https://ui.adsabs.harvard.edu/abs/2021ATel14944....1F} {14944, 1}

\bibitem[\protect\citeauthoryear{{Gaffey}}{{Gaffey}}{1976}]{1976JGR....81..905G}
{Gaffey} M.~J.,  1976, \mn@doi [\jgr] {10.1029/JB081i005p00905}, \href
  {https://ui.adsabs.harvard.edu/abs/1976JGR....81..905G} {81, 905}

\bibitem[\protect\citeauthoryear{{Gaffey}, {Bell}, {Brown}, {Burbine},
  {Piatek}, {Reed}  \& {Chaky}}{{Gaffey} et~al.}{1993}]{1993Icar..106..573G}
{Gaffey} M.~J.,  {Bell} J.~F.,  {Brown} R.~H.,  {Burbine} T.~H.,  {Piatek}
  J.~L.,  {Reed} K.~L.,   {Chaky} D.~A.,  1993, \mn@doi [\icarus]
  {10.1006/icar.1993.1194}, \href
  {https://ui.adsabs.harvard.edu/abs/1993Icar..106..573G} {106, 573}

\bibitem[\protect\citeauthoryear{{Harris}}{{Harris}}{1996}]{1996DDA....27..501H}
{Harris} A.~W.,  1996, in AAS/Division of Dynamical Astronomy Meeting \#27.
  p.~5.01

\bibitem[\protect\citeauthoryear{{Harris} \& {Lupishko}}{{Harris} \&
  {Lupishko}}{1989}]{harris1989}
{Harris} A.~W.,  {Lupishko} D.~F.,  1989, in {Binzel} R.~P.,  {Gehrels} T.,
  {Matthews} M.~S.,  eds, Asteroids II. pp 39--53

\bibitem[\protect\citeauthoryear{{Hergenrother} \& {Whiteley}}{{Hergenrother}
  \& {Whiteley}}{2011}]{2011Icar..214..194H}
{Hergenrother} C.~W.,  {Whiteley} R.~J.,  2011, \mn@doi [\icarus]
  {10.1016/j.icarus.2011.03.023}, \href
  {https://ui.adsabs.harvard.edu/abs/2011Icar..214..194H} {214, 194}

\bibitem[\protect\citeauthoryear{{Holmberg}, {Flynn}  \&
  {Portinari}}{{Holmberg} et~al.}{2006}]{2006MNRAS.367..449H}
{Holmberg} J.,  {Flynn} C.,   {Portinari} L.,  2006, \mn@doi [\mnras]
  {10.1111/j.1365-2966.2005.09832.x}, \href
  {https://ui.adsabs.harvard.edu/abs/2006MNRAS.367..449H} {367, 449}

\bibitem[\protect\citeauthoryear{{Kaasalainen} \& {Torppa}}{{Kaasalainen} \&
  {Torppa}}{2001}]{2001Icar..153...24K}
{Kaasalainen} M.,  {Torppa} J.,  2001, \mn@doi [\icarus]
  {10.1006/icar.2001.6673}, \href
  {https://ui.adsabs.harvard.edu/abs/2001Icar..153...24K} {153, 24}

\bibitem[\protect\citeauthoryear{{Kaasalainen}, {Torppa}  \&
  {Muinonen}}{{Kaasalainen} et~al.}{2001}]{2001Icar..153...37K}
{Kaasalainen} M.,  {Torppa} J.,   {Muinonen} K.,  2001, \mn@doi [\icarus]
  {10.1006/icar.2001.6674}, \href
  {https://ui.adsabs.harvard.edu/abs/2001Icar..153...37K} {153, 37}

\bibitem[\protect\citeauthoryear{{Kole{\'n}czuk}, {Kwiatkowski},
  {Kami{\'n}ska}, {Kami{\'n}ski}, {Colas}, {Klotz}, {Kim}  \&
  {Birlan}}{{Kole{\'n}czuk} et~al.}{2022}]{2022EPSC...16.1161K}
{Kole{\'n}czuk} P.,  {Kwiatkowski} T.,  {Kami{\'n}ska} M.,  {Kami{\'n}ski} K.,
  {Colas} F.,  {Klotz} A.,  {Kim} T.,   {Birlan} M.,  2022, in European
  Planetary Science Congress. pp EPSC2022--1161, \mn@doi{10.5194/epsc2022-1161}

\bibitem[\protect\citeauthoryear{{Lazzarin}, {Marchi}, {Magrin}  \&
  {Licandro}}{{Lazzarin} et~al.}{2005}]{2005MNRAS.359.1575L}
{Lazzarin} M.,  {Marchi} S.,  {Magrin} S.,   {Licandro} J.,  2005, \mn@doi
  [\mnras] {10.1111/j.1365-2966.2005.09006.x}, \href
  {https://ui.adsabs.harvard.edu/abs/2005MNRAS.359.1575L} {359, 1575}

\bibitem[\protect\citeauthoryear{{Mahlke}, {Carry}  \& {Mattei}}{{Mahlke}
  et~al.}{2022}]{2022A&A...665A..26M}
{Mahlke} M.,  {Carry} B.,   {Mattei} P.~A.,  2022, \mn@doi [\aap]
  {10.1051/0004-6361/202243587}, \href
  {https://ui.adsabs.harvard.edu/abs/2022A&A...665A..26M} {665, A26}

\bibitem[\protect\citeauthoryear{{Mainzer} et~al.,}{{Mainzer}
  et~al.}{2011}]{2011ApJ...741...90M}
{Mainzer} A.,  et~al., 2011, \mn@doi [\apj] {10.1088/0004-637X/741/2/90}, \href
  {https://ui.adsabs.harvard.edu/abs/2011ApJ...741...90M} {741, 90}

\bibitem[\protect\citeauthoryear{{Milliken}, {Hiroi}  \&
  {Patterson}}{{Milliken} et~al.}{2016}]{2016LPI....47.2058M}
{Milliken} R.~E.,  {Hiroi} T.,   {Patterson} W.,  2016, in 47th Annual Lunar
  and Planetary Science Conference. Lunar and Planetary Science Conference.
p.~2058

\bibitem[\protect\citeauthoryear{{Mommert}}{{Mommert}}{2017}]{2017A&C....18...47M}
{Mommert} M.,  2017, \mn@doi [Astronomy and Computing]
  {10.1016/j.ascom.2016.11.002}, \href
  {https://ui.adsabs.harvard.edu/abs/2017A&C....18...47M} {18, 47}

\bibitem[\protect\citeauthoryear{{Monteiro}, {Silva}, {Tamayo}, {Rodrigues}  \&
  {Lazzaro}}{{Monteiro} et~al.}{2020}]{2020MNRAS.495.3990M}
{Monteiro} F.,  {Silva} J.~S.,  {Tamayo} F.,  {Rodrigues} T.,   {Lazzaro} D.,
  2020, \mn@doi [\mnras] {10.1093/mnras/staa1401}, \href
  {https://ui.adsabs.harvard.edu/abs/2020MNRAS.495.3990M} {495, 3990}

\bibitem[\protect\citeauthoryear{{Narita} et~al.,}{{Narita}
  et~al.}{2019}]{narita2019}
{Narita} N.,  et~al., 2019, \mn@doi [Journal of Astronomical Telescopes,
  Instruments, and Systems] {10.1117/1.JATIS.5.1.015001}, \href
  {https://ui.adsabs.harvard.edu/abs/2019JATIS...5a5001N} {5, 015001}

\bibitem[\protect\citeauthoryear{{Perna} et~al.,}{{Perna}
  et~al.}{2018}]{2018P&SS..157...82P}
{Perna} D.,  et~al., 2018, \mn@doi [\planss] {10.1016/j.pss.2018.03.008}, \href
  {https://ui.adsabs.harvard.edu/abs/2018P&SS..157...82P} {157, 82}

\bibitem[\protect\citeauthoryear{{Pieters} \& {Hiroi}}{{Pieters} \&
  {Hiroi}}{2004}]{2004LPI....35.1720P}
{Pieters} C.~M.,  {Hiroi} T.,  2004, in {Mackwell} S.,  {Stansbery} E.,  eds,
  Lunar and Planetary Science Conference. Lunar and Planetary Science
  Conference.
p.~1720

\bibitem[\protect\citeauthoryear{{Popescu}, {Birlan}, {Binzel}, {Vernazza},
  {Barucci}, {Nedelcu}, {DeMeo}  \& {Fulchignoni}}{{Popescu}
  et~al.}{2011}]{2011A&A...535A..15P}
{Popescu} M.,  {Birlan} M.,  {Binzel} R.,  {Vernazza} P.,  {Barucci} A.,
  {Nedelcu} D.~A.,  {DeMeo} F.,   {Fulchignoni} M.,  2011, \mn@doi [\aap]
  {10.1051/0004-6361/201117118}, \href
  {https://ui.adsabs.harvard.edu/abs/2011A&A...535A..15P} {535, A15}

\bibitem[\protect\citeauthoryear{{Popescu}, {Birlan}  \& {Nedelcu}}{{Popescu}
  et~al.}{2012}]{2012A&A...544A.130P}
{Popescu} M.,  {Birlan} M.,   {Nedelcu} D.~A.,  2012, \mn@doi [\aap]
  {10.1051/0004-6361/201219584}, \href
  {https://ui.adsabs.harvard.edu/abs/2012A&A...544A.130P} {544, A130}

\bibitem[\protect\citeauthoryear{{Popescu}, {Birlan}, {Nedelcu}, {Vaubaillon}
  \& {Cristescu}}{{Popescu} et~al.}{2014}]{2014A&A...572A.106P}
{Popescu} M.,  {Birlan} M.,  {Nedelcu} D.~A.,  {Vaubaillon} J.,   {Cristescu}
  C.~P.,  2014, \mn@doi [\aap] {10.1051/0004-6361/201424064}, \href
  {https://ui.adsabs.harvard.edu/abs/2014A&A...572A.106P} {572, A106}

\bibitem[\protect\citeauthoryear{{Popescu}, {Licandro}, {Carvano}, {Stoicescu},
  {de Le{\'o}n}, {Morate}, {Boac{\u{a}}}  \& {Cristescu}}{{Popescu}
  et~al.}{2018}]{2018A&A...617A..12P}
{Popescu} M.,  {Licandro} J.,  {Carvano} J.~M.,  {Stoicescu} R.,  {de Le{\'o}n}
  J.,  {Morate} D.,  {Boac{\u{a}}} I.~L.,   {Cristescu} C.~P.,  2018, \mn@doi
  [\aap] {10.1051/0004-6361/201833023}, \href
  {https://ui.adsabs.harvard.edu/abs/2018A&A...617A..12P} {617, A12}

\bibitem[\protect\citeauthoryear{{Popescu} et~al.,}{{Popescu}
  et~al.}{2019}]{2019A&A...627A.124P}
{Popescu} M.,  et~al., 2019, \mn@doi [\aap] {10.1051/0004-6361/201935006},
  \href {https://ui.adsabs.harvard.edu/abs/2019A&A...627A.124P} {627, A124}

\bibitem[\protect\citeauthoryear{{Pravec} \& {Harris}}{{Pravec} \&
  {Harris}}{2000}]{2000Icar..148...12P}
{Pravec} P.,  {Harris} A.~W.,  2000, \mn@doi [\icarus]
  {10.1006/icar.2000.6482}, \href
  {https://ui.adsabs.harvard.edu/abs/2000Icar..148...12P} {148, 12}

\bibitem[\protect\citeauthoryear{{Rubincam}}{{Rubincam}}{2000}]{2000Icar..148....2R}
{Rubincam} D.~P.,  2000, \mn@doi [\icarus] {10.1006/icar.2000.6485}, \href
  {https://ui.adsabs.harvard.edu/abs/2000Icar..148....2R} {148, 2}

\bibitem[\protect\citeauthoryear{{S{\'a}nchez} \& {Scheeres}}{{S{\'a}nchez} \&
  {Scheeres}}{2014}]{2014M&PS...49..788S}
{S{\'a}nchez} P.,  {Scheeres} D.~J.,  2014, \mn@doi [Meteorit. Planet. Sci.]
  {10.1111/maps.12293}, \href
  {https://ui.adsabs.harvard.edu/abs/2014M&PS...49..788S} {49, 788}

\bibitem[\protect\citeauthoryear{{Shepard} et~al.,}{{Shepard}
  et~al.}{2008}]{2008Icar..195..184S}
{Shepard} M.~K.,  et~al., 2008, \mn@doi [\icarus]
  {10.1016/j.icarus.2007.11.032}, \href
  {https://ui.adsabs.harvard.edu/abs/2008Icar..195..184S} {195, 184}

\bibitem[\protect\citeauthoryear{{Simion}, {Popescu}, {Licandro}, {Vaduvescu},
  {de Le{\'o}n}  \& {Gherase}}{{Simion} et~al.}{2021}]{2021MNRAS.508.1128S}
{Simion} N.~G.,  {Popescu} M.,  {Licandro} J.,  {Vaduvescu} O.,  {de Le{\'o}n}
  J.,   {Gherase} R.~M.,  2021, \mn@doi [\mnras] {10.1093/mnras/stab2561},
  \href {https://ui.adsabs.harvard.edu/abs/2021MNRAS.508.1128S} {508, 1128}

\bibitem[\protect\citeauthoryear{{Tatsumi}, {Tinaut-Ruano}, {de Le{\'o}n},
  {Popescu}  \& {Licandro}}{{Tatsumi} et~al.}{2022}]{2022A&A...664A.107T}
{Tatsumi} E.,  {Tinaut-Ruano} F.,  {de Le{\'o}n} J.,  {Popescu} M.,
  {Licandro} J.,  2022, \mn@doi [\aap] {10.1051/0004-6361/202243806}, \href
  {https://ui.adsabs.harvard.edu/abs/2022A&A...664A.107T} {664, A107}

\bibitem[\protect\citeauthoryear{Tedesco, Tholen  \& Zellner}{Tedesco
  et~al.}{1982}]{tedesco1982eight}
Tedesco E.~F.,  Tholen D.~J.,   Zellner B.,  1982, The Astronomical Journal,
  87, 1585

\bibitem[\protect\citeauthoryear{{Tholen}}{{Tholen}}{1984}]{1984PhDT.........3T}
{Tholen} D.~J.,  1984, PhD thesis, University of Arizona

\bibitem[\protect\citeauthoryear{{Tody}}{{Tody}}{1986}]{1986SPIE..627..733T}
{Tody} D.,  1986, in {Crawford} D.~L.,  ed.,  Society of Photo-Optical
  Instrumentation Engineers (SPIE) Conference Series Vol. 627, Instrumentation
  in astronomy VI. p.~733, \mn@doi{10.1117/12.968154}

\bibitem[\protect\citeauthoryear{{Warner}, {Harris}  \& {Pravec}}{{Warner}
  et~al.}{2009}]{2009Icar..202..134W}
{Warner} B.~D.,  {Harris} A.~W.,   {Pravec} P.,  2009, \mn@doi [\icarus]
  {10.1016/j.icarus.2009.02.003}, \href
  {https://ui.adsabs.harvard.edu/abs/2009Icar..202..134W} {202, 134}

\bibitem[\protect\citeauthoryear{{Zappala}, {Cellino}, {Barucci}, {Fulchignoni}
   \& {Lupishko}}{{Zappala} et~al.}{1990}]{1990A&A...231..548Z}
{Zappala} V.,  {Cellino} A.,  {Barucci} A.~M.,  {Fulchignoni} M.,   {Lupishko}
  D.~F.,  1990, \aap, \href
  {https://ui.adsabs.harvard.edu/abs/1990A&A...231..548Z} {231, 548}

\bibitem[\protect\citeauthoryear{{de Le{\'o}n}, {Licandro}, {Serra-Ricart},
  {Pinilla-Alonso}  \& {Campins}}{{de Le{\'o}n}
  et~al.}{2010}]{2010A&A...517A..23D}
{de Le{\'o}n} J.,  {Licandro} J.,  {Serra-Ricart} M.,  {Pinilla-Alonso} N.,
  {Campins} H.,  2010, \mn@doi [\aap] {10.1051/0004-6361/200913852}, \href
  {https://ui.adsabs.harvard.edu/abs/2010A&A...517A..23D} {517, A23}

\bibitem[\protect\citeauthoryear{de Mello, Da~Silva, Da~Silva  \&
  De~Nader}{de~Mello et~al.}{2014}]{de2014photometric}
de Mello G.~P.,  Da~Silva R.,  Da~Silva L.,   De~Nader R.,  2014, Astronomy \&
  Astrophysics, 563, A52

\makeatother
\end{thebibliography}








\bsp	
\label{lastpage}
\end{document}